\documentclass[prl,twocolumn,showpacs,amsmath,amssymb,superscriptaddress]{revtex4}

\usepackage[dvips]{graphicx}

\usepackage{graphics}

\usepackage{dcolumn}

\usepackage{bm}

\usepackage[usenames]{color}

\begin{document}

\title{Anisotropic magnetoresistance components in (Ga,Mn)As}


\author{A.~W.~Rushforth}


\affiliation{School of Physics and Astronomy, University of Nottingham, Nottingham NG7 2RD, UK}

\author{K.~V\'yborn\'y}

\affiliation{Institute of Physics ASCR, Cukrovarnick\'a 10, 162 53 Praha 6, Czech Republic}

\author{C.~S.~King}

\affiliation{School of Physics and Astronomy, University of Nottingham, Nottingham NG7 2RD, UK}
\author{K.~W.~Edmonds}
\affiliation{School of Physics and Astronomy, University of Nottingham, Nottingham NG7 2RD, UK}
\author{R.~P.~Campion}

\affiliation{School of Physics and Astronomy, University of Nottingham, Nottingham NG7 2RD, UK}

\author{C.~T.~Foxon}

\affiliation{School of Physics and Astronomy, University of Nottingham, Nottingham NG7 2RD, UK}

\author{J.~Wunderlich}

\affiliation{Hitachi Cambridge Laboratory, Cambridge CB3 0HE, UK}

\author{A.~C.~Irvine}

\affiliation{Microelectronics Research Centre, Cavendish Laboratory, University of Cambridge, CB3 0HE, UK}

\author{P.~Va\v{s}ek}

\affiliation{Institute of Physics ASCR, Cukrovarnick\'a 10, 162 53 Praha 6, Czech Republic}

\author{V.~Nov\'ak}

\affiliation{Institute of Physics ASCR, Cukrovarnick\'a 10, 162 53 Praha 6, Czech Republic}

\author{K.~Olejn\'{\i}k}

\affiliation{Institute of Physics ASCR, Cukrovarnick\'a 10, 162 53 Praha 6, Czech Republic}

\author{Jairo~Sinova}
\affiliation{Department of Physics, Texas A\&M University, College
Station, TX 77843-4242, USA}

\author{T.~Jungwirth}

\affiliation{Institute of Physics ASCR, Cukrovarnick\'a 10, 162 53 Praha 6, Czech Republic}

\affiliation{School of Physics and Astronomy, University of Nottingham, Nottingham NG7 2RD, UK}

\author{B.~L.~Gallagher}

\affiliation{School of Physics and Astronomy, University of Nottingham, Nottingham NG7 2RD, UK}

\date{\today}

\begin{abstract}
Our experimental and theoretical study of  the non-crystalline and crystalline
components of the anisotropic magnetoresistance (AMR) in (Ga,Mn)As is aimed at exploring basic physical aspects
of this relativistic transport effect.
The non-crystalline AMR
reflects anisotropic lifetimes of the  
holes due to  
polarized  Mn impurities while  the crystalline AMR is associated with valence band warping.
We find that the sign of the non-crystalline  AMR is  determined
by the form of spin-orbit coupling in the host band and
by the relative strengths of the non-magnetic and magnetic contributions to the impurity potential.
We develop experimental methods directly yielding the non-crystalline and crystalline AMR components which
are then independently analyzed. We
report the observation of an AMR dominated by a large uniaxial crystalline component
and show that AMR  can be modified by local strain relaxation.
We discuss generic implications of our experimental and theoretical findings including predictions for
non-crystalline AMR sign reversals in dilute moment systems.
\end{abstract}

\pacs{75.47.-m, 75.50.Pp, 75.70.Ak}

\maketitle

Anisotropic magnetoresistance (AMR) is a response of
carriers in magnetic materials to changes of the
magnetization orientation.
Despite its 
importance in magnetic recording technologies the understanding of the microscopic
physics of this spin-orbit (SO) coupling induced
effect is relatively poor.
Phenomenologically, AMR has a non-crystalline component, arising from the lower symmetry for a specific
current
direction, and crystalline components arising from the crystal symmetries \cite{Doring:1938_a,vanGorkom:2001_a}.
In ferromagnetic metals, values for these coefficients can be obtained by 
numerical {\em ab initio} transport calculations \cite{Banhart:1995_a}, but these have no clear connection to the standard physical model of
transport arising from spin dependent scattering of current carrying low mass $s$-states into heavy-mass $d$-states \cite{McGuire:1975_a}.
Experimentally, the non-crystalline and, the
typically much weaker, crystalline AMR  components in metals have been indirectly extracted  from fitting the total
AMR angular dependences \cite{vanGorkom:2001_a}.

Among the remarkable AMR features of (Ga,Mn)As ferromagnetic semiconductors are the
opposite sign of the non-crystalline component (compared to most
metal ferromagnets) and the  crystalline terms reflecting the
rich magnetocrystalline anisotropies \cite{Baxter:2002_a,Jungwirth:2003_b,Tang:2003_a,Matsukura:2004_a,Goennenwein:2004_a,Wang:2005_c,Limmer:2006_a}.
Microscopic numerical simulations \cite{Jungwirth:2002_c,Jungwirth:2003_b}
consistently describe the sign and magnitudes of the non-crystalline AMR and 
capture the more
subtle crystalline terms associated with {\em e.g.} growth-induced strain \cite{Jungwirth:2002_c,Matsukura:2004_a}.
As in metals, however, the basic microscopic physics of the AMR
still needs to be elucidated which is the aim of the work presented here.


Theoretically, we separate the non-crystalline and crystalline components by turning off and on band warping and
match
numerical microscopic simulations with model analytical results. This provides the physical
interpretation of the origin of AMR, and of the sign of the non-crystalline term
in particular.
Experimentally, we obtain direct and independent access to the non-crystalline and
crystalline AMR components using Hall bars fabricated along the principle crystalline axes and Corbino disk samples.
The
method is first established in the standard
(Ga,Mn)As films before discussing the unique behavior we observe in ultra-thin low-conductive
(Ga,Mn)As layers, which show a large, uniaxial crystalline component dominated AMR. Finally we
demonstrate  how crystalline AMR components can be strongly modified by local strain relaxation~\cite{Wunderlich:relax}.

The experimental data presented in this paper were measured in compressively strained 25nm and 5nm Ga$_{0.95}$Mn$_{0.05}$As films grown by low temperature molecular beam epitaxy on  GaAs [001] substrates. Optical lithography was used to fabricate Hall bars aligned along the [100], [010], [110] and $[1\bar{1}0]$  directions, of width
45$\mu$m with voltage probes separated by 285$\mu$m  and Corbino disks of inner diameter 800$\mu$m and outer diameter 1400$\mu$m
in which current flows radially in the plane of the material.
Electron beam lithography was used to fabricate  1$\mu$m wide Hall bars in a 25nm Ga$_{0.95}$Mn$_{0.05}$As film. All magnetoresistances were measured with the magnetization in the plane of the device, {\em i.e.}, in the pure AMR geometry with zero (antisymmetric) Hall signal.

The phenomenological decomposition of the AMR of (Ga,Mn)As into various terms allowed by symmetry is obtained by
extending the standard phenomenology \cite{Doring:1938_a},
to systems with cubic [100] plus uniaxial [110] anisotropy. With this we write the longitudinal AMR as, $\Delta\rho_{xx}/\rho_{av}
= C_I\cos2\phi + C_U\cos2\psi + C_C\cos4\psi
+ C_{I,C}\cos(4\psi-2\phi)$,
where $\Delta\rho_{xx}=\rho_{xx}-\rho_{av}$, $\rho_{av}$ is the  $\rho_{xx}$ averaged over 360$^{o}$ in the plane of the film, $\phi$ is the angle between the magnetization unit vector ${\bf \hat{M}}$ and the current ${\bf I}$,
and $\psi$ the angle between  ${\bf \hat{M}}$ and the [110] crystal direction.
The four contributions are the non-crystalline term, the lowest order uniaxial and cubic crystalline terms, and a crossed non-crystalline/crystalline term.
The purely crystalline terms are excluded by symmetry for the transverse AMR and we obtain,
$\Delta\rho_{xy}/\rho_{av}=C_I\sin2\phi - C_{I,C}\sin(4\psi-2\phi)$.

Microscopically we explain the emergence of the AMR components starting from the
valence-band kinetic-exchange description of (Ga,Mn)As 
with metallic conductivities, which is an established qualitative
and often semiquantitative theoretical approach \cite{Jungwirth:2006_a,Jungwirth:2007_a}. The
description is based on the canonical Schrieffer-Wolff
transformation of the Anderson Hamiltonian
which for (Ga,Mn)As replaces hybridization of Mn $d$-orbitals with As and Ga $sp$-orbitals by an effective
spin-spin interaction of  $L=0,S=5/2$
local-moments with host valence band states. These states,
which carry all the SO-coupling,
can be described by the ${\bf k}\cdot{\bf p}$
Kohn-Luttinger Hamiltonian \cite{Dietl:2001_b,Jungwirth:2006_a}.

\begin{figure}[ht]
\vspace*{-0cm}

%
\hspace*{-0cm}\includegraphics[width=1\columnwidth,angle=-0]{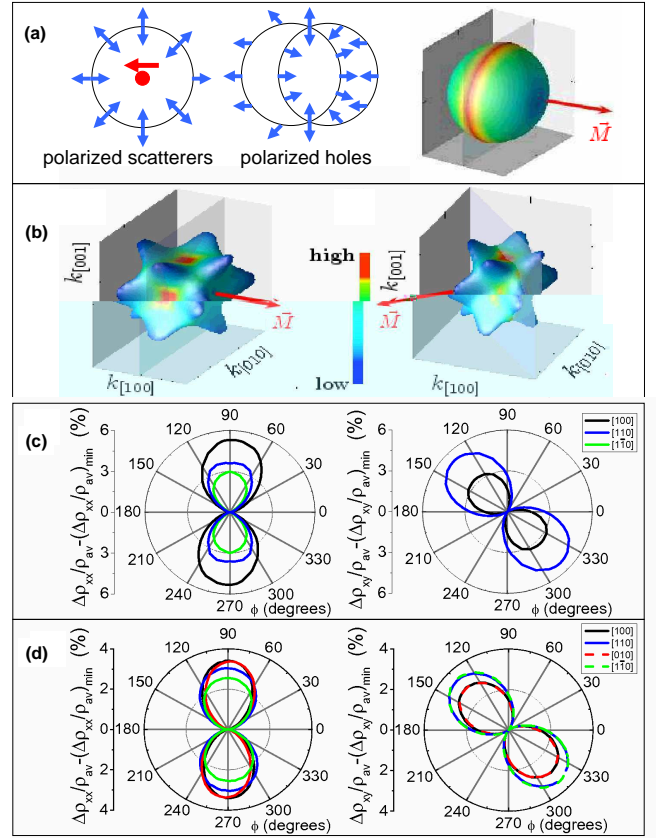}

\vspace*{-0cm}
\caption{(a) Non-crystalline AMR in spherical bands: 2D Cartoons of AMR mechanisms and calculated anisotropic
scattering rate on
the 3D Fermi surface of the minority heavy-hole band in  Ga$_{0.95}$Mn$_{0.05}$As. (b) Non-crystalline and
crystalline AMR on warped bands: calculated anisotropic
scattering rates for ${\bf \hat{M}}\parallel$[100] and [110] axes. (c) Calculated and (d) measured (at 4.2~K) longitudinal and transverse AMR for Ga$_{0.95}$Mn$_{0.05}$As as a function of the angle between
${\bf \hat{M}}$~and~${\bf I}$.
The legend shows the direction of the current. The y-axes show $\Delta\rho/\rho_{av}$ shifted such that the minimum is at zero.}
\label{f1}
\end{figure}

In these dilute moment systems
 two distinct microscopic mechanisms lead to anisotropic
carrier lifetimes, as illustrated in Fig.~\ref{f1}(a):
One combines the SO-coupling in the carrier band with polarization of randomly distributed magnetic scatterers and the
other
with polarization of the carrier band itself resulting in an asymmetric band-spin-texture.
Although acting simultaneously in real systems, theoretically we can turn both
mechanisms on and off independently. We find
that the former mechanism clearly dominates in (Ga,Mn)As which allows us to neglect spin-splitting of the valence band
in the following qualitative discussion.
This is  further simplified by focusing on the non-crystalline AMR in the heavy-hole Fermi
surfaces  in the spherical, ${\bf  s}\parallel {\bf k}$, spin-texture approximation \cite{Jungwirth:2002_a}
(see Fig.~\ref{f1}(a))
and considering scattering off a $\delta$-function potential $\propto(\alpha+ {\bf \hat{M}}\cdot {\bf s})$.
Here ${\bf  s}$ and  ${\bf k}$ are
the carrier spin-operator and wavevector, and
$\alpha$ represents the ratio of non-magnetic and magnetic parts of the impurity potential. Assuming
a proportionality between
conductivity and lifetimes of carriers with ${\bf k}||{\bf I}$ we
obtain,
\begin{equation}
\frac{\sigma({\bf \hat{M}}\parallel {\bf I})}{\sigma({\bf \hat{M}}\perp {\bf I})}=
\left(\alpha^2+\frac14\right)\left(\alpha^2+\frac{1}{12}\right)\left(\alpha^2-\frac14\right)^{-2} \;.
\label{eq1}
\end{equation}
Therefore when
$\alpha\ll 1$,
one expects
$\sigma({\bf \hat{M}}\parallel I)<\sigma({\bf \hat{M}}\perp I)$  (as is usually observed in metallic ferromagnets).
But the sign of the non-crystalline AMR reverses at a
relatively weak non-magnetic potential ($\alpha=1/\sqrt{20}$ in the model), its magnitude is then
maximized when the two terms are comparable
($\alpha=1/2$), and, for this mechanism, it vanishes when the magnetic term is
much weaker than the non-magnetic term ($\alpha\rightarrow\infty$).

Physically, carriers moving along ${\bf \hat{M}}$, {\em i.e.} with ${\bf  s}$ parallel or antiparallel to ${\bf \hat{M}}$,
experience the strongest
scattering potential among all Fermi surface states when $\alpha=0$, giving
$\sigma({\bf \hat{M}}\parallel I)<\sigma({\bf \hat{M}}\perp I)$. When
the non-magnetic potential is present, however, it can more efficiently cancel the magnetic term for carriers moving
along   ${\bf \hat{M}}$, and for relatively small $\alpha$ the sign of AMR flips. Since $\alpha<1/\sqrt{20}$
is unrealistic for the
magnetic acceptor Mn in GaAs \cite{Jungwirth:2002_c,Jungwirth:2006_a} we obtain $\sigma({\bf \hat{M}}\parallel I)>\sigma({\bf \hat{M}}\perp I)$, consistent with experiment.
Our analysis also predicts that when the SO-coupling in the host band is of the form ${\bf  s}\perp {\bf k}$,
as in the  Rashba-type 2D systems, or when Mn forms an isovalent pure magnetic impurity, {\em e.g.}  in  II-VI semiconductors, the sign of the non-crystalline AMR will be reversed.

Numerical simulations  of hole scattering rates, illustrated in Fig.~1(a) on a color-coded minority heavy-hole Fermi surface,
were obtained within the spherical approximation but including the hole spin polarization, light-hole and split-off
valence bands, and realistic non-magnetic and magnetic Mn impurity potentials \cite{Jungwirth:2002_c}.
The simulations confirm the
qualitative validity of the analytical, non-crystalline AMR expressions of
Eqs.~(\ref{eq1}).
The
additional crystalline AMR terms are obtained when the spherical approximation is relaxed and band warping is included in
the Kohn-Luttinger Hamiltonian \cite{Dietl:2001_b,Jungwirth:2006_a}.
The enhanced scattering of holes moving perpendicular to $\bf \hat{M}$ seen in Fig.1(b)
reflects the persistence of a strong non-crystalline component in the
warped bands. The presence of the crystalline components in these anisotropic Fermi surfaces
is also clearly apparent in Fig.1(b).
In (Ga,Mn)As, the crystalline
terms
reflect the biaxial cubic anisotropy  of these zincblende compounds combined with a [110] uniaxial  component \cite{note1}.

We conclude the theory discussion by showing in Fig.1(c) full numerical
Boltzmann theory simulations 
of the AMR for a weakly (15\%) compensated
Ga$_{0.95}$Mn$_{0.05}$As material. The non-crystalline AMR contributes strongly and, as explained above,
leads to a higher resistance state for  ${\bf I}\perp{\bf \hat{M}}$. Differences among AMRs for current along the [100], [110], and $[1\bar{1}0]$ directions show that cubic and uniaxial crystalline terms are also sizable. This phenomenology is systematically observed in experimental AMRs of weakly or moderately compensated metallic (Ga,Mn)As films. Typical data for such systems, represented by the 25nm Ga$_{0.95}$Mn$_{0.05}$As film with 3.6\% AMR, are shown in Fig.1(d) for the Hall bars patterned along the [100], [010], [110], and $[1\bar{1}0]$ directions. In these measurements a saturating magnetic field of 1T is applied in the plane of the film and the magnetization vector follows the external field direction.

Above we have shown that
diluted magnetic semiconductors like (Ga,Mn)As allow us to formulate the theory of AMR from
the understanding of the very basic microscopic mechanisms.
In what follows we focus on several unique experimental aspects of the AMR in these systems. The high crystalline quality and metallic character of the samples allow us to produce low contact resistant Hall bars accurately orientated along the principle crystallographic axes, from which it is possible to extract the independent contributions to the AMR. We are also able to fabricate low contact resistance Corbino disk samples for which the averaging over the radial current lines eliminates all effects originating from a specific direction of the current. Corbino measurements are possible in these materials because they are near perfect single crystals but with low carrier density and mobility (compared with single crystal metals) and so can have source-drain resistances large compared with the contact resistances.

Measured results for a Corbino device fabricated from the same 25nm Ga$_{0.95}$Mn$_{0.05}$As film as used for the Hall bars are shown in Fig. 2(a). The AMR signal is an order of magnitude weaker than in the Hall bars and is clearly composed of a uniaxial and a cubic contribution.
Fig.~2(a) also shows the crystalline components of the AMR extracted by fitting the Hall bar data to the phenomenological
longitudinal and transverse AMR expressions \cite{Corbino}.
Fig.~2(b) shows the consistency for the coefficients C$_{I,C}$, C$_{U}$ and C$_{C}$
when extracted from the Hall bar and Corbino disk data over the whole range up to the Curie temperature (80K).
Note that the uniaxial crystalline term, C$_{U}$, becomes the dominant term for T$\geqslant$30K. This correlates with the uniaxial component of the magnetic anisotropy which dominates for T$\geqslant$30K as observed by SQUID magnetometry measurements (not shown). Our work shows that in (Ga,Mn)As ferromagnets, the symmetry breaking mechanism behind the previously reported \cite{Sawicki:2004_a} uniaxial magneto-crystalline anisotropy in the magnetization  also contributes to {the AMR}.
\begin{figure}[ht]

\hspace*{0cm}\includegraphics[width=.37\columnwidth,angle=-90]{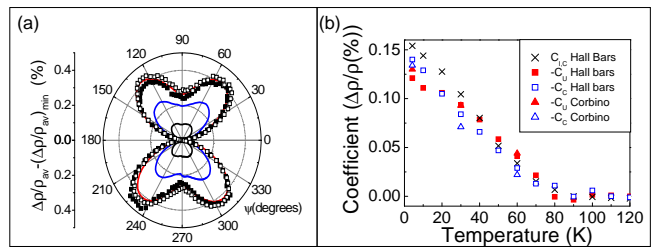}
\caption{(a) AMR of the 25nm Ga$_{0.95}$Mn$_{0.05}$As film in the Corbino geometry at 4.2~K (red line), 30~K (blue line), and 60~K (black line) and the crystalline component extracted from the Hall bars (AMR[110]+AMR[1$\bar{1}$0])/2 (closed points) and (AMR[100]+AMR[010])/2 (open points). (b) Temperature dependence of the crystalline terms extracted from the Hall bars and Corbino devices.}
\label{f2}
\end{figure}

We now discuss the unique AMR phenomenology observed on ultra thin (5nm) Ga$_{0.95}$Mn$_{0.05}$As films. Measurements on the Hall bars in Fig. 3(a) show that the AMR is very different from that observed in the 25nm film. Application of the phenomenological analysis to the Hall bar data shows that this behavior is a consequence of the crystalline terms dominating the AMR with the uniaxial crystalline term being the largest. SQUID magnetometry on 5nm Ga$_{0.95}$Mn$_{0.05}$As films consistently shows that the uniaxial component of the magnetic anisotropy dominates over the whole temperature range \cite{Rushforth:2006_a}. The Corbino disk AMR data for a nominally identical 5nm film, shown in Fig. 3(b), confirm our observation of the highly unconventional 6\% AMR  totally dominated by the uniaxial crystalline term.

The 5nm films have lower Curie temperatures ($T_{C}\approx$30K) than the 25nm films and become highly resistive at low temperature indicating that they are close to the metal-insulator transition. (The 25nm films show metallic behavior
upto the lowest measured temperatures.)
The strength of the effect in the 5~nm films is remarkable and it is not captured by theory simulations assuming
weakly disordered, fully delocalized (Ga,Mn)As valence bands. It might be related to the expectation that magnetic interactions become more anisotropic with increasing localization of the holes near their parent Mn ions
as the metal-insulator transition is approached~\cite{Jungwirth:2006_a}.

\begin{figure}[ht]

\hspace*{0cm}\includegraphics[width=.7\columnwidth,angle=-90]{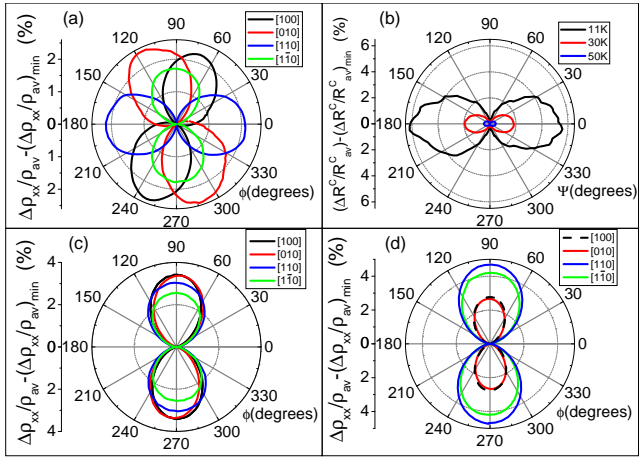}
\caption{Longitudinal AMR of the 5nm Ga$_{0.95}$Mn$_{0.05}$As Hall bars. T=20K. (b) AMR of a 5nm Ga$_{0.95}$Mn$_{0.05}$As film in the Corbino geometry. (c) AMR for macroscopic Hall bars and (d) narrow (1$\mu$m wide) Hall bars.}
\label{f3}
\end{figure}

Finally, we demonstrate how the crystalline terms can be tuned by the use of lithographic patterning to induce an additional uniaxial anisotropy in very narrow Hall bars. In recent studies \cite{Wunderlich:relax} it has been found that the patterning allows the in-plane compressive strain in the (Ga,Mn)As film to relax in the direction along the width of the Hall bar and this can lead to an additional uniaxial component in the magnetocrystalline
anisotropy for bars with widths on the order of 1$\mu$m or smaller. Figs.~3(c),(d) show the AMR of 45$\mu$m wide bars and 1$\mu$m wide bars fabricated from nominally identical
25nm Ga$_{0.95}$Mn$_{0.05}$As wafers. For the 45$\mu$m bars, the cubic crystalline symmetry leads to the AMR along [100] and [010] being larger than along [110] and [1$\bar{1}$0]. For the narrow bars we observe the opposite relationship. This is consistent with the addition of an extra uniaxial component, whose
presence in the magnetocrystalline anisotropy is confirmed by SQUID magnetometry measurements (not shown),
which adds 0.8\% to the AMR when current is along [110] and [1$\bar{1}$0] and subtracts 0.4\% when the current is along [100] and [010]. These post-growth lithography induced modifications are significant fractions of the total AMR
of the parent (Ga,Mn)As material.

To conclude, we have described the non-crystalline AMR in (Ga,Mn)As as a combined effect of the SO-coupled spin-texture in
the host band and polarized scatterers containing  non-magnetic and magnetic impurity potentials. The additional
crystalline terms are associated with band warping effects which reflect the underlying crystal symmetry. Our theory
should apply to a large family of
related dilute moment systems suggesting, {\em e.g.}, that the non-crystalline
AMR sign flips when the
${\bf  s}\parallel {\bf k}$ Kohn-Luttinger spin-texture is replaced by the ${\bf  s}\perp {\bf k}$ Rashba-type
SO-coupling of asymmetric 2D systems, or when Mn forms an isovalent, pure magnetic impurity as is the case in  II-VI
semiconductor structures. On the experimental side we have established a technique for direct measurement
of the crystalline AMR components by
utilizing the Corbino disk geometry, which should also be applicable to other
systems which combine high crystalline quality with relatively high resistivity. We report that in (Ga,Mn)As, ultra-thin films can be epitaxially grown with
AMR dominated by a large uniaxial crystalline component and that in standard films the crystalline
components can be modified by microscale lithography induced lattice relaxations.

We acknowledge collaborations with J. Chauhan, D. Taylor, K.Y. Wang, and M. Sawicki, and support from  EU Grant  IST-015728, from UK Grant GR/S81407/01, from CR  Grants 202/05/0575, 202/04/1519, FON/06/E002, AV0Z1010052, and LC510, from ONR Grant N000140610122, and from SWAN. J. Sinova is a Cottrell Scholar of Research Corporation.


\begin{thebibliography}{18}
\bibitem{Corbino}Strictly, in a Corbino geometry we measure $G^{C}=I/V=(\sigma_{xx}$+$\sigma_{yy})/2$ and plot $R^{C}=1/G^{C}$. If we define $R^{C}=1/G^{C}$ and note that $\rho_{xy}\ll\rho_{xx}$ then it is possible to show that $\Delta R^{C}/R_{av}^{C}\approx(\Delta R_{[110]}+\Delta R_{[1\bar{1}0]})/R_{av}$ from the Hall bars = $C_{U}cos2\psi+C_{C}cos4\psi$.
\end{thebibliography}

\end{document}